\documentclass{article}
\usepackage{amssymb}
\usepackage{indentfirst} 
\usepackage{graphicx}
\usepackage{verbatim}
\usepackage{geometry}
\usepackage{graphicx, caption, subcaption}
\usepackage{floatrow}
\usepackage{colortbl}
\usepackage{soul}
\usepackage{xcolor}
\usepackage{algorithmic}
\usepackage{algorithm}
\usepackage{ulem}
\usepackage{multirow}
\usepackage{makecell}
\usepackage{amsmath}
\usepackage[hyphens]{url}
\usepackage{wasysym}
\usepackage{stfloats}
\date{}
\title{Crash Report Data Analysis for Creating Scenario-Wise, Spatio-Temporal Attention Guidance to Support Computer Vision-based Perception of Fatal Crash Risks}

\author{%
Yu Li, Muhammad Monjurul Karim, Ruwen Qin\footnote{Corresponding author, ruwen.qin@stonybrook.edu, 2424 Computer Science Building, Stony Brook University, Stony Brook, NY 11794-4424}, Zuhui Wang, Zhaozheng Yin,  Zeyi Sun}

\begin{document}
\maketitle 
\section{Abstract} 
Reducing traffic fatalities and serious injuries is a top priority of the US Department of Transportation. With the rapid development of sensor and Artificial Intelligence (AI) technologies, the computer vision (CV)-based crash anticipation in the near-crash phase is receiving growing attention. The ability to perceive fatal crash risks earlier is also critical because it will improve the reliability of crash anticipation. Yet, annotated image data for training a reliable AI model for the early visual perception of crash risks are not abundant. The Fatality Analysis Reporting System (FARS) contains big data of fatal crashes. It is a reliable data source for learning the relationship between driving scene characteristics and fatal crashes to compensate for the limitation of CV. Therefore, this paper develops a data analytics model, named scenario-wise, Spatio-temporal attention guidance, from fatal crash report data, which can estimate the relevance of detected objects to fatal crashes from their environment and context information. First, the paper identifies five sparse variables that allow for decomposing the 5-year (2013-2017) fatal crash dataset to develop scenario-wise attention guidance. Then, exploratory analysis of location- and time-related variables of the crash report data suggests reducing fatal crashes to spatially defined groups. The group's temporal pattern is an indicator of the similarity of fatal crashes in the group. Hierarchical clustering and K-means clustering merge the spatially defined groups into six clusters according to the similarity of their temporal patterns. After that, association rule mining discovers the statistical relationship between the temporal information of driving scenes with crash features, such as the first harmful event and the manner of collisions, for each cluster. The paper shows how the developed attention guidance supports the design and implementation of a preliminary CV model that can identify objects of a possibility to involve in fatal crashes from their environment and context information.

\hfill\break%
\noindent\textit{Keywords}: Crash Data Analysis , Association Rule Mining , Clustering , Spatio-Temporal Information , Attention Guidance , Visual Perception

\section{Introduction}
\label{Introduction}
Enhancing traffic safety is a top priority of the US Department of Transportation (USDOT). In 2018, there were 36,560 deaths in motor vehicle crashes \cite{FARS}. Federal Highway Administration (FHWA) aims to reduce motor vehicle fatalities and serious injuries across the transportation system \cite{zerodeathvision}. Various safety enhancement programs consider a multi-pronged approach that encompasses countermeasures of enforcement, education, emergency response, and engineering (named 4E) \cite{HSIP4ES2016}. 4E's countermeasures are usually developed for specific problems, such as pedestrian-involved crashes, work zone-related crashes, large truck-involved crashes, and teenage drivers. State DOTs, local entities, and other transportation agencies are active in implementing those safety countermeasures.

With the rapid development of sensor technologies and artificial intelligence, Computer Vision (CV) is receiving growing attention in the transportation industry, for instance, the use of CV to assist both automated driving vehicles and human drivers (e.g.,  \cite{velez2016embedding, yao2019egocentric}). Recently,  crash-related problems such as learning crash-related features automatically for crash anticipation adopt the CV technique (e.g., \cite{bao2020uncertainty, suzuki2018anticipating, chan2016anticipating}). CV is helpful for crash anticipation in the near-crash phase (e.g., 2$\sim$4 seconds before crashes) due to its ability to detect and track objects and to learn crash-related features of them automatically. The ability to identify crash risk earlier, for instance, a half minutes before crashes, is also desired because it allows for allocating attention to relevant agents to leverage up the performance of the near-crash phase analysis. Yet, in stand-alone, CV is not sufficient to reliably anticipate crash risks when crash-related features have not been distinct. For example, a pedestrian in a non-crossing area at night is more likely to involve in a fatal crash than one who is crossing the road at a controlled intersection. CV has the ability to detect the presence of a pedestrian in the driving scene, locate the pedestrian, and tell the environment. Yet, CV lacks the ability to determine how likely the driving scene would associate with certain types of fatal crashes. After a big dataset of pre-crash images becomes available, CV can also build the capability for early visual perception of fatal crash risks. Before that, an alternative solution is to use other sources of data to compensate for the current limitation of CV.   

Crash report systems, such as the Fatality Analysis Reporting System (FARS) \cite{FARS} and the Crash Report Sampling System (CRSS) \cite{CRSS}, are publicly available datasets that provide comprehensive descriptions of crashes from various perspectives. FARS is a suitable data source for learning the relationship between driving scene characteristics and fatal crashes. Studies of fatal crashes have widely used FARS \cite{clark2002predicted, briggs2005fatality, yasmin2015pooling}. Analyses of crash report data provide insights into safety enhancement (e.g., \cite{SafetydataInitiative, SolvingSafety, TRIP, data2017knowledge}) and support the safety strategy development and law enforcement (e.g., \cite{abdel2011study, harb2008freeway, pour2016investigating, sharma2016s}). Not only fatal crash causes (e.g., bad weather) but driving scene characteristics that have an association with crashes (e.g., rush hours) can indicate fatal crash risks. Although widely used in crash data analysis, basic descriptive statistics and graphic techniques are insufficient for discovering complex relationships among variables. Researchers have started using advanced data analytics methods to study crash report data.

This paper is motivated to develop a data analytics model, named attention guidance. The object of this paper is to analyze the features and variable distribution of crash data to develop the guidance by the chance of the crash consequence, instead of classification studies. Therefore, the fatal crash report data retrieved from FARS is the data that adopted in this paper. The attention guidance supports the CV-based early perception of fatal crash risks. Such a CV model can recognize driving scene characteristics and, accordingly, determine how likely the driving scene is associated with certain types of fatal crashes. Consequently, attentive vehicles, pedestrians, and other agents that are likely to involve in fatal crashes are known. Based on our best knowledge, no work has been done to timely meet this need. Comprehensive attention guidance requires systematic development effort. To cover every detail within one research paper is difficult. Due to the following reasons, this paper chooses to focus on fatal crashes in the ``common scenario" that is most commonly confronted by human drivers and automated driving systems. First, fatal crashes are the most concerning consequence of an average economic cost of \$1.7 million per death in 2018 \cite{Costs_Injuries}. Preventing fatal crashes is an urgent need. Second, the literature shows that fatal crash risks have varied features or causes in different driving scenarios, necessitating the scenario-wise crash risk analysis. Third, the CV-based driving sense analysis can classify driving scenes by scenarios and identify scene characteristics to retrieve the corresponding attention guidance. Last but not least, the method developed and illustrated in this study applies to other scenarios.

The remainder of this paper is organized as follows. The next section summarizes the related work. Then, the proposed approach to developing the attention guidance is explained in section 3. Section 4 discusses the implementation detail. In the end, section 5 draws the conclusion from this study and recommends future work worth further exploration.

\section{Literature Review}
\label{literature}

The literature of this study are in two streams: crash risk indicators and methods of exploratory data analysis. The literature is briefly discussed below.

\subsection{Crash Risk Indicators}
Crash risk factors are reviewed and evaluated from various perspectives. For example, \cite{papadimitriou2019review} summarized 59 infrastructure-related crash risk factors that can be structured in a hierarchical taxonomy of 10 areas, such as road functional class, adverse road environment, presence of work zones, cross-section deficiencies, and junction deficiencies. This study also developed a method to rank the factors as Risky, Probably Risky, Probably not Risky, or Unclear. Besides providing clear guidance for infrastructure design and improvement, this review evidences the presence of a relationship that links crash risks to weather, work zones, and other above-mentioned infrastructure-related factors. Motivated by the increasing availability of real-time traffic data and the need for proactive safety management, \cite{theofilatos2014review} performed a comprehensive literature review to determine the effect of traffic and weather characteristics on road safety. Traffic characteristics reviewed in this paper include volume, density, and speed, and weather characteristics include precipitation, low visibility, wind, and air temperature. This review emphasizes the joint impact of traffic and weather characteristics. Haddon matrix is another common approach to classifying risk factors, which groups factors by the pre-crash, crash, and post-crash phases, concerning humans, vehicles, and the environmental condition \cite{WHORiskFactors}. Haddon matrix also suggests a systems approach to crash risk analysis, which examines the interactions of the above-mentioned multiple components. 

While existing review papers on risk factors provided useful information to this study, reviews dedicated to the driving scene characteristics for featuring fatal crash risks are not found. Those characteristics are related to time, location, and environment. The first two categories are mainly context information relevant to fatal crashes, whereas the third category is contributing causes for fatal crashes. 

Crash risks may be inferred from time-related characteristics of driving scenes, such as hour, week, month, and season because fatal crashes are found to distribute non-homogeneously in the temporal space. \cite{pour2016investigating} found the volume of severe injury cases from 20:00 to 5:59 is higher than other hours. \cite{chong2018epidemiology} split the 24 hours of the day into four segments and sorted them in the increasing order of pedestrian-involved fatal cases. The dashboard of pedestrian crash data on the DOT website showed that the volume of pedestrian-involved crashes in wintertime is higher than in other seasons  \cite{PedestrianTrafficFatalities}. More crashes happened between 17:00 and 19:59 during the wintertime, but after 20:00 in other seasons. Compared to weekdays, more crashes occurred after midnight on Saturday and Sunday. 

Similarly, crash risks may be inferred from location-related characteristics of driving scenes, such as road type and intersection type. \cite{PedestrianTrafficFatalities} showed the volume of pedestrian-involved crashes on other principal arterial is the highest. \cite{rovvsek2017identifying} found that injury severity in the circumstance of inappropriate speed is most likely to be fatal on the freeway, regional, or local roads. Statistical comparisons indicate work zone activities are riskier on the interstate than on other road types \cite{liu2016role}. \cite{meng2011evaluation} found rear-end crash risk is higher on expressways than on other arterial. Pedestrian fatalities in non-intersection areas are almost three times of those at intersections \cite{PedestrianTrafficFatalities}. 

The environmental condition influences drivers' ability to operate vehicles. Visibility is one explored widely (e.g., \cite{ pour2016investigating, harb2008freeway, rovvsek2017identifying, peng2017assessing}). Darkness, poor lighting conditions, and no lighting conditions are factors contributing to the severity of pedestrian-involved crashes  \cite{pour2016investigating}. Pedestrian fatalities happened in the dark condition were three times more than in the daylight condition \cite{PedestrianTrafficFatalities}. Bad weather increases the chance of crashes. \cite{wang2015real} found that the wet pavement surface condition increases the crash risk by 77\% in the weaving segment of the expressway. \cite{abdel2011study} identified high-chance scenarios of crashes in fog, smoke fog, and smoke conditions. A bad weather condition impacts the injury severity more than a poor lighting or road surface condition does \cite{rovvsek2017identifying}. The environmental condition also influences humans' behavior. People tend to be less patient and more likely to violate traffic rules in undesirable weather conditions \cite{zhai2019diagnostic}, and drivers become more aggressive in dealing with challenging roadway and weather conditions  \cite{hamdar2016weather}.

\subsection{Exploratory Analysis of Crash Report Data}
Data exploration along with visualization is an analytics method to summarize and understand data. Typical graphical techniques used in THE exploratory analysis of crash data include scatter charts, line charts, bar or column charts, bubble charts, and heat maps \cite{kamruzzaman2019exploratory, abdel2011study, harb2008freeway, smith2016spring, chong2018epidemiology, delmelle2012spatial}. Spatial and temporal patterns of crashes are useful decision-making tools. Moran's I is a measurement to evaluate the existence of a pattern of clustering by calculating the spatial autocorrelation \cite{moran1948interpretation}. Pearson's r is a coefficient that can measure the correlation between two patterns. \cite{lee2001developing} integrated Pearson's r and Moran's I to develop a bivariate spatial association measure, named L. Data mining techniques such as clustering and association rule mining are also proposed for discovering new knowledge or patterns, particularly from big crash data. \cite{rovvsek2017identifying} used a classification tree to find key risk factors impacting the severity of traffic injuries on Slovenian roads. \cite{montella2011data} innovatively converted results of classification trees into association rules. Each terminal node of the tree is treated as the consequence of a rule and all the splits of parent nodes are  the antecedent.

\section{The Approach}
\label{Approach}
\begin{figure*}[htbp]
    \centering
    \includegraphics[width=\linewidth]{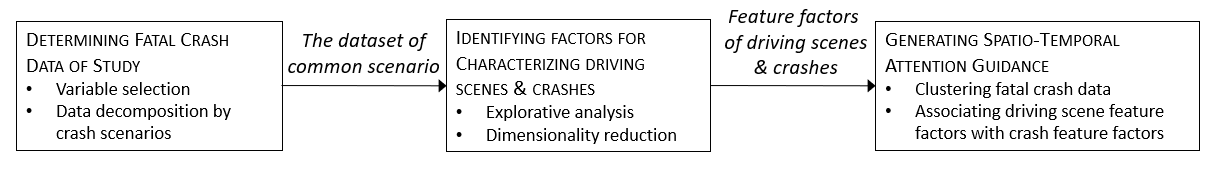}
    \caption{The proposed approach to creating the spatio-temporal attention guidance}
    \label{fig:Approach}
\end{figure*}

The proposed approach to creating the scenario-wise, spatio-temporal attention guidance is a three-step process described in Figure \ref{fig:Approach}. Details of the approach are discussed below.

\subsection{Determining Fatal Crash Data of Study}
This study used the 2013-2017 U.S. fatal crash data retrieved from FARS \cite{FARS}. There are 162,104 fatal crashes during this five-year period. FARS has more than 40 variables about fatal crashes  \cite{FARSManual}.

\subsubsection{Variable Selection}
This study chose thirteen variables that fall in the following four categories to characterize driving scenes, which can be either recognized by CV methods (e.g., classification, segmentation, and detection) or provided by systems equipped in vehicles (e.g., the clock). 
\begin{itemize}
\vspace{-0.8em}
    \item Time-related: ``Hour of Crash", ``Day of Week", ``Month of Crash";\vspace{-0.8em}
    \item Location-related:  ``Functional System", ``Relation to Trafficway", ``Relation to Junction-Specific Location", ``Related to Junction-Within Interchange Area", ``Type of Intersection";\vspace{-0.8em}
    \item ``Environment-related: ``Light Condition", ``Atmospheric Conditions";\vspace{-0.8em}
    \item Special crash types: ``Work Zone", ``School Bus Related", ``Related Factor-Crash Level".
\vspace{-0.8em}
\end{itemize}
The study further chose two additional variables to describe crashes, which are "First Harmful Event" (FHE) and ``Manner of Collision" (MC). FHE describes the first injury or damage producing event of the crash. MC identifies the manner in which two motor vehicles in transport initially came together without regard to the direction of the force. MC is considered for crashes where the FHE is ``Motor vehicle in transport”.

\subsubsection{Data Decomposition for Generating Scenarios}
Among the thirteen variables for characterizing driving scenes, five are relevant to a small portion of fatal crashes. They are ``School Bus Related", ``Work Zone Related", ``Within Interchange Area", ``Related Factor-Crash Level", and ``Atmospheric Conditions". That is, if the five variables are coded as binary variables, most values of these binary variables are zeros. The data sparsity of them provides an opportunity to decompose the dataset into a finite number of mutually exclusive and collectively exhaustive subsets. Fatal crashes in each subset are identical on one or several variables for characterizing driving scenes. This study defines each subset as a {\it scenario}. For example, all the work zone related fatal crashes can be defined as a crash scenario. A driving scene is mapped to one and only one crash scenario. A hypothesis in this study is that individual scenarios have their unique crash risk patterns that are more distinct than the pattern of the population.

\begin{figure*}[htb]
    \centering
    \includegraphics[width=0.8\linewidth]{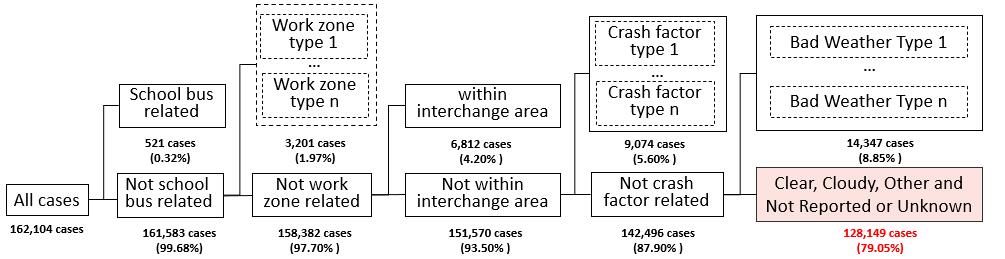}
    \caption{Data decomposition by crash scenarios that defined in accordance to five sparse variables}
    \label{fig:Low_Chance}
\end{figure*}

This study proposed a hierarchical tree method to decompose the fatal crash data into mutually exclusive and collectively inclusive subsets with each corresponding to a crash scenario, as Figure \ref{fig:Low_Chance} illustrates. The study ranked the five variables for classifying fatal crash scenarios in the increasing order of non-zero elements and decomposed the dataset in that sequence. The number of school bus related fatal crashes is only 521 (0.32\% of the 162,104 cases), smaller than the number of any other type of crashes. Therefore, the first step of data decomposition was made according to the variable ``School Bus Related". After that, crashes not related to school bus were further split according to the variable ``Work Zone". This process was continuing until all of the five variables have been used for the data decomposition. Figure \ref{fig:Low_Chance} shows that the biggest subset is the {\it common scenario} composed of fatal crashes not related to school bus, work zones, interchange area, crash factors, and bad weather. This subset contains 128,149 fatal crashes, equal to  79.05\% of total cases during the five-year period. The common scenario dataset was analyzed in the remainder of this paper with respect to the remaining eight variables for characterizing driving scenes and the two variables for featuring crashes.

\subsection{Identifying Factors for Characterizing Driving Scenes}
Among the eight variables for characterizing driving scenes, three are time-related, four are location-related, and one is environment-related. The environment-related factor is the "Light Condition", which is verified to strongly correlate with ``Hour of Crash". A basic exploratory analysis of the common scenario dataset was performed to develop an initial understanding of it on the time and location dimensions, respectively.

\subsubsection{Fatal Crash Distribution on Time-related Variables} 
\begin{figure*}[hbt]
\centering \includegraphics[width=0.65\linewidth]{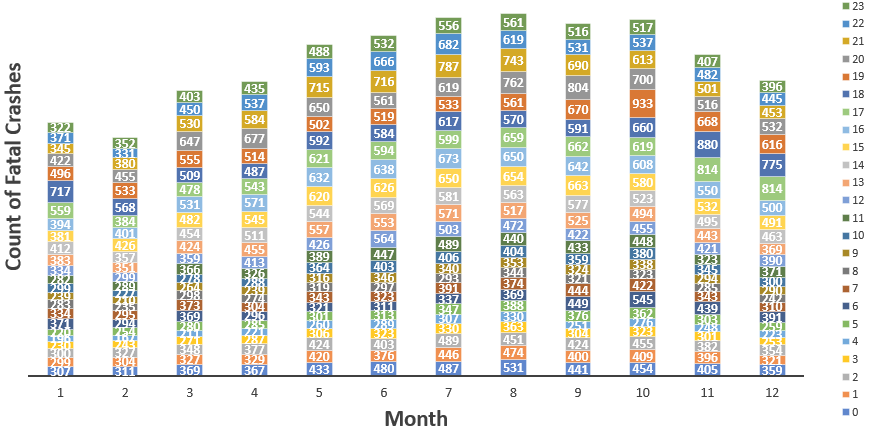}
 \caption{The frequency distribution of fatal crash (2012-2017) on months of the year, split by hours of the day}
\label{fig:month}
\end{figure*}
Figure \ref{fig:month} counts the fatal crashes from the common scenario dataset by month and by hour. The distributions of crashes by month and by hour are not even. The volume of fatal crashes is the smallest in February, increases afterward, reaches the highest level in fall, and drops gradually in winter. Afternoon and evening hours (13:00-23:59) have more fatal crashes than other hours of the day. Moreover, peak hours of fatal crashes demonstrate a seasonal pattern. Peak hours are 18:00-19:59 in February, 19:00-20:59 in March, 20:00-21:59 in April to September, 19:00-20:59 in October, and 17:00-18:59 in winter.

\begin{figure*}[htb]
\centering
\begin{subfigure}[b]{.62\textwidth}
\centering
\includegraphics[width=\textwidth]{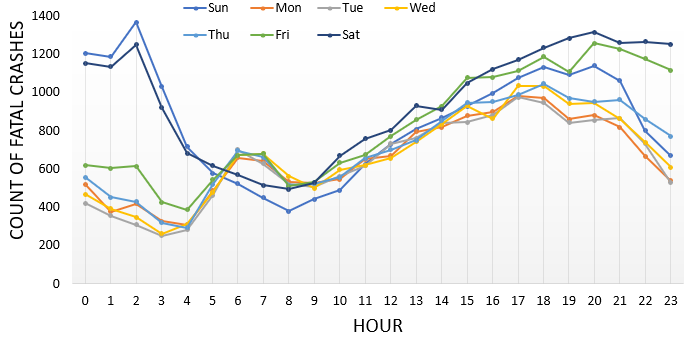}
\caption{Hourly series of fatal crashes on seven days of the week}
\end{subfigure}
\begin{subfigure}[b]{.31\textwidth}
\centering
\includegraphics[width=0.6\textwidth]{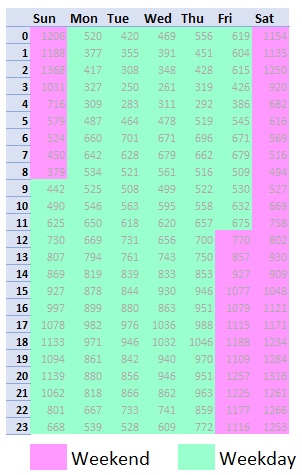} 
\caption{Weekday and weekend split}
\end{subfigure}
\caption{Reduction of seven daily patterns to the weekday and weekend Patterns} \label{fig:hour_by_Week}
\end{figure*}

Figure \ref{fig:hour_by_Week}(a) further displays the trend lines of fatal crashes in each day of the week. The trend lines of Monday, Tuesday, Wednesday, and Thursday are close to each other, suggesting the existence of a weekday pattern. The trend line of Saturday is separated from the weekday pattern, representing a weekend pattern. The trend lines of Friday and Sunday stretch to both patterns. The afternoon to the evening of Friday is close to the weekend pattern, and so is Sunday from midnight to the early morning. Observations from Figures \ref{fig:month} and  \ref{fig:hour_by_Week}(a) indicate the presence of two month-hour 2D distributions of fatal crashes, one for weekdays and the other for the weekends.

Splitting the crash data into the subsets of weekdays and weekends is a k-means clustering problem solved as integer optimization. Let $n_i^j$ represent the number of fatal crashes in hour $i$ of week $j$, for $i=0,1,\dots,23$ and $j\in\{Mo, Tu, We, Th, Fr, Sa, Su\}$. Let $x_i^D$ and $X_i^E$ stand for the average number of fatal crashes at hour $i$ of weekdays and weekends, respectively. They are initialized as 
    \begin{equation}
    \label{eq:XijD}
     x_{i}^{D}=\left(n_{i}^{Mo}+n_{i}^{Tu}+n_{i}^{We}+n_{i}^{Th}\right)/4, \;x_{i}^{E}=n_{i}^{Sa}.
     \end{equation}
The hourly series of fatal crashes on Friday is split into two segments; that is, $\{n_i^{Fr}|i=0,\dots,23\}=\{n_i^{Fr}|i=0,\dots,p\}\cup \{n_i^{Fr}|i=p+1,\dots,23\}$. The first segment of Friday belongs to the weekday and the second segment is the weekend. The place to split the time series needs to be chosen so that the sum of squared errors due to merging the two segments of Friday data into the weekend and weekend time series, respectively, is minimized:
   \begin{equation}
   \label{eq:SS_Fr}
 \min_{p\in \{0,\dots,23\}}   ss^{Fr} (p)= \sum_{i=0}^{p}\left(n_{i}^{Fr}-x_{i}^{D}\right)^{2}+\sum_{i=p+1}^{23}\left(n_{i}^{Fr}-x_{i}^{E}\right)^{2}.
    \end{equation}
Similarly, the hourly series of fatal crashes of Sunday is split as two segments. The first $q$ hours belong to the weekend and the remaining hours belong to the weekday.

The counts of crashes,
$\{n_i^j|\forall\; i, j\}$, in Eqs. (\ref{eq:XijD}-\ref{eq:SS_Fr}) were estimated using the 128,149 fatal crashes. The optimal values of $q$ an $p$ were found to be 11 and 8, respectively, and Figure \ref{fig:hour_by_Week}(b) displays the result. That is, 9:00 of Sunday to 11:59 of Friday is the time frame for the weekday crash pattern and 12:00 of Friday to 8:59 of Sunday is the time frame for the weekend pattern.
    
\subsubsection{Fatal Crash Distribution on Location-related Variables}
Fatal crashes at locations of the same characteristics may share certain similarities. Road type and road configuration straightforwardly characterize the location of a driving scene. ``Functional System" (FUNC\_SYS) is the variable describing the road type, and the road configuration is described by the combination of another three location-related variables: ``Relation to Trafficway"(REL\_ROAD), ``Relation to Junction-Specific Location" (RELJCT2), and ``Type of Intersection" (TYP\_INT). Due to their hierarchical relationship, this study merged TYP\_INT into RELJCT2 to create a new variable JUN\_INT by replacing the attribute ``Intersection or related" of RELJCT2 with the specific intersection types such as ``Four-Way Intersection",  ``T-Intersection", ``Y-Intersection", and so on. In the remainder of the paper, REL\_ROAD and JUN\_INT jointly define the road configuration. Table \ref{tab:Four_Clusters}(b) summarizes the attributes of these variables.

\begin{figure*}[htb]
        \centering
        \includegraphics[width=\linewidth]{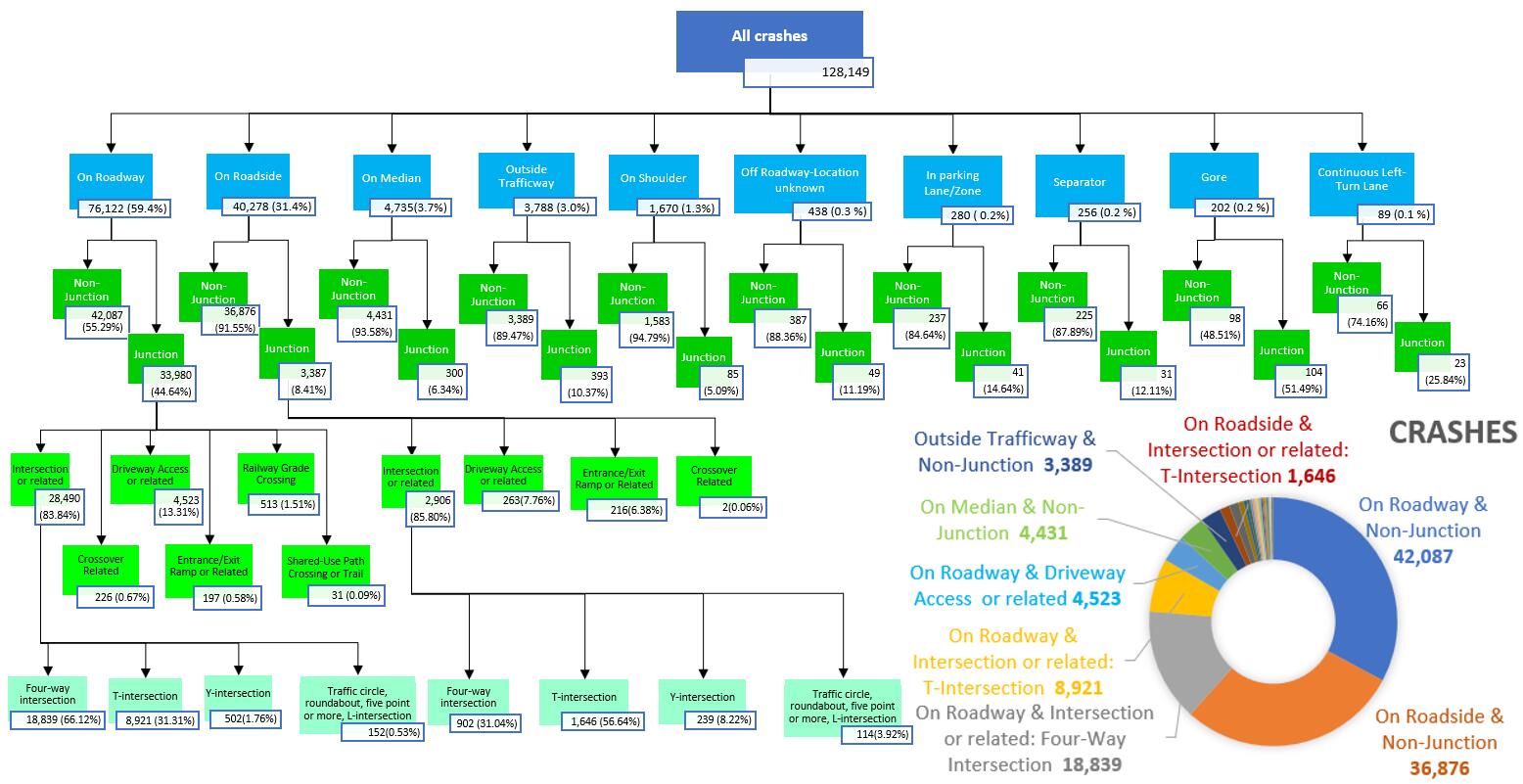}
        \caption{Fatal crash distribution on the road configuration tree}
        \label{fig:Location_tree}
\end{figure*}
Figure \ref{fig:Location_tree} is a tree structure that shows the distribution of fatal crashes on the different areas of roads. Over 85\% fatal crashes occurred at the following five areas: on the roadway and non-junction area (32.8\%), on the roadside and non-junction area (28.8\%), on the roadway and four-way intersections (14.7\%), on the roadway and T-intersections (7.0\%), and on the roadway and driveway access or related (3.5\%). Although this distribution is partially correlated with the proportions of these areas in the roadway system, users of the system usually have different maneuvers, actions, and behavior at different areas.

\begin{figure*}[hbtp]
    \centering
    \includegraphics[width=0.85\linewidth]{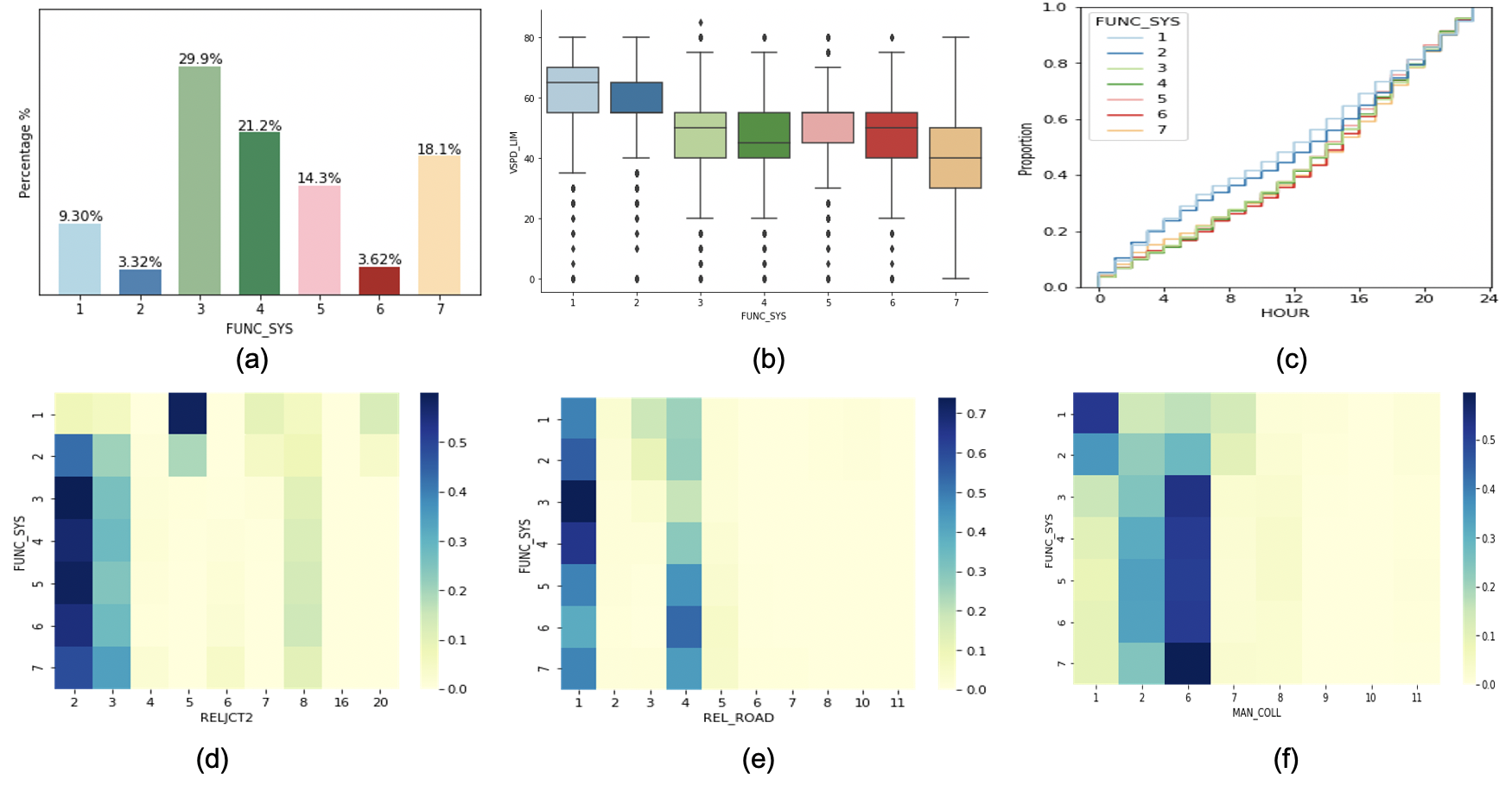}
    \caption{The functional system with seven road types: (a) fatal crash distribution, (b) speed limits, (c) cumulative distribution on hours of the day, (d) distribution on  junction-related locations, (e) distribution on areas of trafficway, (f) distribution on manners of collisions.}
    \label{fig:Fig5}
\end{figure*}
The functional system includes seven road types listed in Table \ref{tab:Four_Clusters}(b). Among the seven types, interstate (1) has the most mobility, whereas local (7) has the most accessibility. The seven road types serve different purposes, which may influence the fatal crash patterns on them. The six plots (a-f) in Figure \ref{fig:Fig5} are some observations.
\begin{itemize}
\vspace{-0.8em}
    \item [(a)] is the distribution of fatal crashes across the seven road types. Other principal arterial (3) has the largest volume, about 30\% of total cases, whereas freeways \& expressways (2) has the smallest volume, about 3\%.  \vspace{-0.8em}
    \item [(b)] displays the box plots of the speed limit on the seven road types. The speed limits of interstate (1) and freeways \& expressways (2) are the highest, and the speed limit of local (7) is the lowest. Other arterial (3\&4) and collectors (5\&6) have a similar speed limit, statistically. \vspace{-0.8em}
    \item [(c)] shows the cumulative distribution functions of fatal crashes on hours of the day, split by road types. More cases on interstate (1) and freeways \& expressways (2) happened in the early hours of the day than on other road types.\vspace{-0.8em}
    \item [(d)] is a heat map that examines the distribution of fatal crashes at junction-related locations on the seven road types. The top three locations on interstate (1) are ``5-Entrance/Exit Ramp Related”, ``20-Entrance/Exit Ramp”, and ``7-Crossover Related". The top three locations on freeways \& expressways (2) are ``2-Intersection”, ``3-Intersection Related” and ``5-Entrance/Exit Ramp Related”. The top three locations on other road types are ``2-Intersection”, ``3-Intersection Related” and ``8-Driveway Access Related”.\vspace{-0.8em}
    \item [(e)] is a heatmap that shows the distribution of fatal crashes at different areas of the trafficway on the seven road types. For all the road types, the areas with the largest number of fatal crashes are “1-On Roadway” and “4-On Roadside”. “3-Median” is another area on interstate (1) and freeways \& expressways (2) where some fatal crashes occurred. \vspace{-0.8em}
    \item[(f)] is the distribution of fatal crashes on the Manner of Collision (MC) at each road type. Fatal crashes with the ``1-Front-to-Rear" MC on the interstate (1), and those with the ``6-Angle" MC on Local (7), Collectors (5\&6), and other Arterial (3\&4), are dominantly higher than crashes with other MCs and on other road types.\vspace{-0.8em}
\end{itemize}

\subsection{Developing the Spatio-Temporal Attention Guidance}
Grouping fatal crashes by their location (defined by road type and road configuration) effectively reduce the size of the dataset from 128,149 cases to 184 spatially-defined groups. A hypothesis supports this approach, which says fatal crashes occurred on the same type of roads and at the same area of roads have certain similarities. The exploratory analysis above supports this hypothesis and this study further developed a method to merge those groups into larger groups. If the larger groups are distinct sufficiently, each should have its own attention guidance. 

\subsubsection{Spatially Defined Groups of Fatal Crashes}
The common scenario dataset is broken down into 184 mutually exclusive and collectively exhaustive groups that each contains fatal crashes occurred on the same type of roads and at the same area of roads. For example, the spatially defined group ``Local, On Roadside, and Driveway Access or Related" comprises fatal crashes with FUNC$\_$SYS=7, REL\_ROAD=4, and JUN$\_$INT=4.
\begin{figure*}[htb]
    \centering
    \begin{subfigure}[b]{.48\textwidth}
      \includegraphics[width=\textwidth]{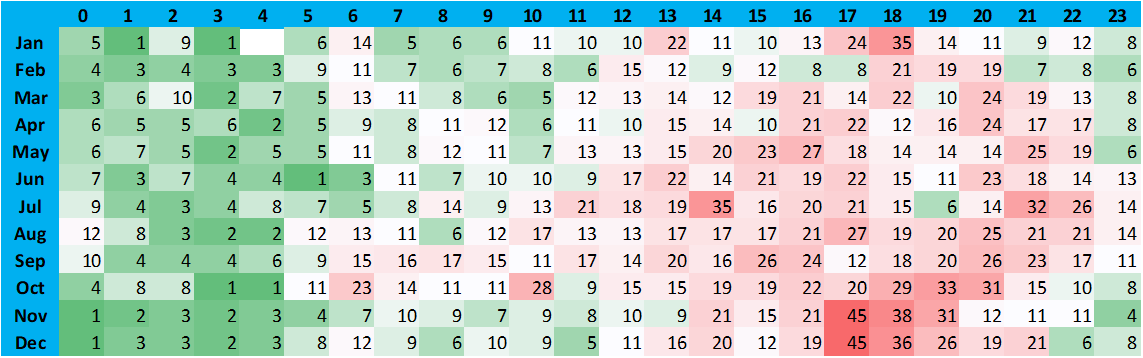}
      \vspace{0.02cm}
      \includegraphics[width=\textwidth]{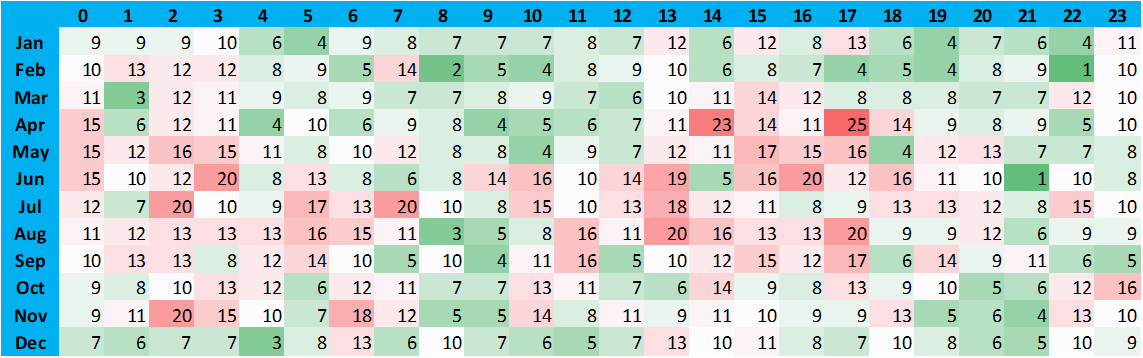}
      \caption {Examples of 2D distribution on the month-hour grid}
    \end{subfigure}
      \hspace{0.4cm}  
    \begin{subfigure}[b]{.4\textwidth}
      \includegraphics[width=\textwidth]{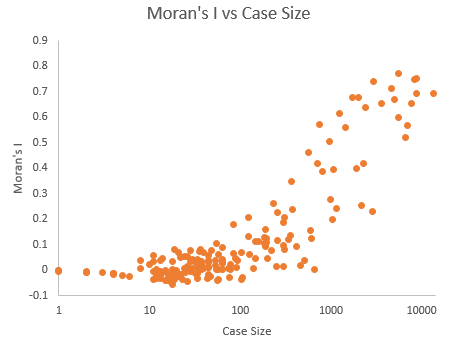}
      \caption{Moran's I vs group size}
    \end{subfigure}
    \caption{Spatially defined groups of fatal crashes}
    \label{fig:moranI_casesize}
\end{figure*}

Then, the month-hour 2D temporal distribution of fatal crashes is created for each of the spatially defined groups. The 2D distribution counts the number of fatal crashes at each spot of the month-hour grid, and it uses colors to visualize the variation of cases on this 12$\times$24 grid - the darker the red (and green) color is, the larger (and less) the volume of cases at that spot of the temporal grid. Figure \ref{fig:moranI_casesize}(a) illustrates two examples. The one on the top is the group ``Principal Arterial – Other, on Roadway, and T-Intersection" with 3,609 fatal crashes. The one at the bottom is the group ``Interstate, On Roadside, and Non-Junction" with 2,858 cases. These two distributions are very different. The distribution on the top has a pattern with a red spot on the right and a green spot on the left. The distribution at the bottom is more random, without a clear pattern. The month-hour 2D temporal pattern is one evidence that cases within a sptailly defined group have similarities. Moran's I was calculated for each spatially defined group to quantify the strength of its temporal pattern. Moran's I in this paper is the autocorrelation of cases at the 288 month-hour spots of the temporal grid:
\begin{equation}
I=\frac{288}{\sum_{i,j=1}^{288}w_{i,j}} \frac{\sum_{i,j=1}^{288}w_{i,j}(y_{i}-\bar{y})(y_{j}-\bar{y})}{\sum_{i=1}^{288}(y_{i}-\bar{y})^{2}},
\label{eq:MoranI}
\end{equation}
where $i$ and $j$ are linear indices of any two spots of the grid. $y_i$ and $y_j$  are the numbers of fatal crashes at spots $i$ and $j$, respectively. $\bar{y}$ is the average number of fatal crashes per spot. $w_{i,j}$ is a binary coefficient that takes one when spots $i$ and $j$ are neighbors and zero otherwise. Each spot has at least three and up to eight neighbors. The value of Moran's I is from -1 to 1, where -1 means perfect dispersion, 0 is perfect randomness, and 1 indicates perfect clustering. 

In Figure \ref{fig:moranI_casesize}(a), the Moran's I value of the distribution on the top is 0.65 and the value of the bottom one is 0.23. Figure \ref{fig:moranI_casesize}(b) further displays the scatter plot of the 184 spatially defined groups with their Moran's I value and the group size. This figure confirms that groups of smaller sizes (e.g., no greater than 100 fatal crashes) are likely to have a random temporal distribution. Groups of larger sizes tend to have clustering patterns, but groups with  similar sizes may have varied strengths in their temporal patterns. Therefore, this study further analyzes the similarity of the 68 groups that each contains more than 100 fatal crashes. There are 123,782 fatal crashes in those 68 groups, which count for 96.6\% of the total 128,149 crashes in the common scenario dataset. Groups of size 100 or less contain 4,367 cases, only 4.4\% of the data in the common scenario dataset.

\subsubsection{Clustering Spatially Defined Groups}

\begin{figure*}[htb]
    \centering
    \begin{subfigure}[b]{.4\textwidth}
      \includegraphics[width=\textwidth]{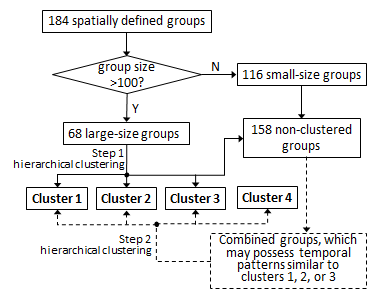}
      \caption{The approach to clustering spatially defined groups}
    \end{subfigure}
      \hspace{0.1cm}  
    \begin{subfigure}[b]{.57\textwidth}
      \includegraphics[width=\textwidth]{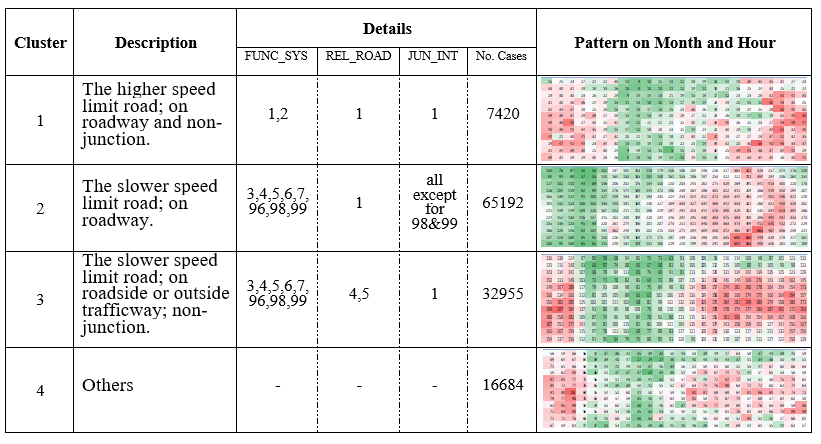}
      \caption{Four spatially defined clusters\protect}
    \end{subfigure}
    \caption{The approach to clustering fatal crashes and the result}
    \label{fig:clustering}
\end{figure*}

The 184 spatially defined groups were clustered in a two-step approach, as Figure \ref{fig:clustering}(a) shows. During the first step, a down-to-top hierarchical clustering algorithm was used to cluster the 68 groups of size greater than 100. The algorithm begins from the bottom of the hierarchical tree, which has 68 single-group clusters (i.e., each cluster contains only one spatially defined group). In each iteration, the Pearson's r is calculated for every pair of clusters and, accordingly, the most similar clusters are merged as a new cluster. The number of clusters is reduced by at least one in each iteration. The process will be terminated automatically when all groups are merged as a single cluster, reaching the top of the tree. But this study observed the growth and formation of three meaningful clusters during the iterative process of clustering and thus terminated the process. The three groups are summarized below and Table \ref{tab:Four_Clusters} lists the detail:
\begin{itemize}
\vspace{-0.8em}
    \item Cluster 1: 2 groups of roads with higher speed limits, on roadway, and non-junction areas; \vspace{-0.8em}
    \item Cluster 2: 18 groups of roads with slower speed limits, on roadway, and non-junction or some types of junction;\vspace{-0.8em}
    \item Cluster 3: 6 groups of the slower speed limit roads, on roadside, and non-junction areas. \vspace{-0.8em}
\end{itemize}

\begin{table*}[htb]   
    \caption{Clustering Results}        
    \label{tab:Four_Clusters}   
    \begin{subtable}[h]{0.35\textwidth}
        \centering
        \caption{Details of the four spatially defined clusters}        
     \includegraphics[scale=0.51]{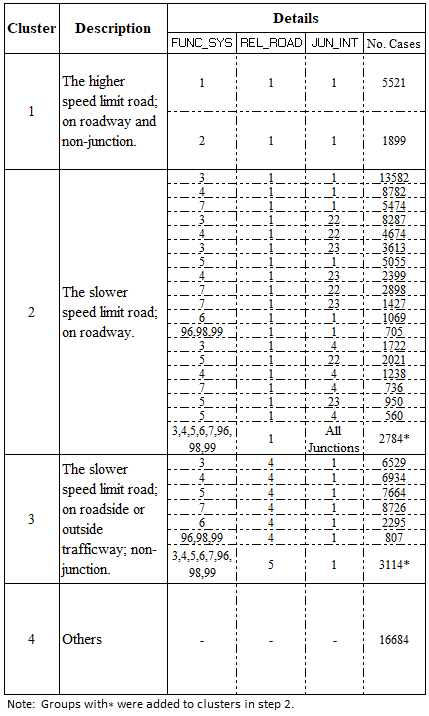}
    \end{subtable}
\hskip 1.8cm
    \begin{subtable}[h]{0.5\textwidth}
        \centering
        \caption{Code and definitions}               \includegraphics[scale=0.65]{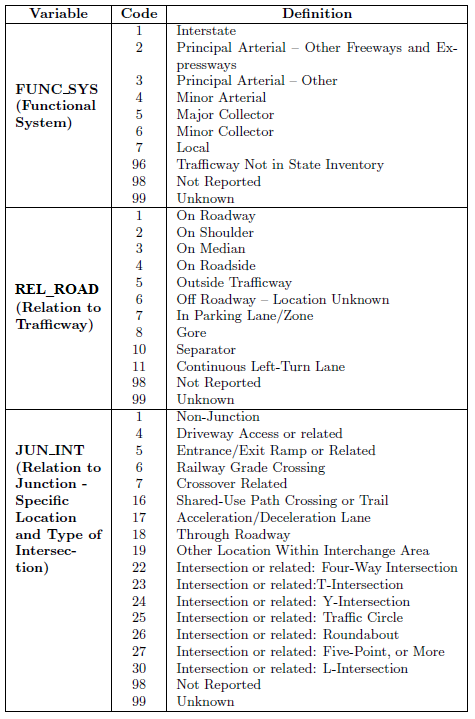}
    \end{subtable}
\end{table*}
When the first step of the clustering process was terminated, 28 out of the 68 groups (85.3\% fatal crashes in the 68 groups) were clustered into the three groups above, and 42 groups still remained as single-group clusters.

The second step of the clustering analysis was focused on the remaining 22,582 fatal crashes in the 158 single-group groups, including the 116 groups of size no greater than 100 and the 42 groups left from the first step of clustering. Some small size groups, if combined into a bigger one, may possess a temporal pattern similar to one of the three clusters initially created from step 1. From observing the three clusters, this study identified a few candidates and successfully merged 2,784 cases into cluster 2, which are located at ``roads with lower speed limits, on the roadway, and other junction types in addition to those already included in cluster 2". Then, 3,114 cases were merged with cluster 3, which are  located at ``roads with lower speed limits, outside trafficway, and non-junction". The hierarchical clustering in the second step was stopped after two iterations. The remaining 16,684 fatal crashes were just treated as cluster 4. These four clusters are depicted in Figure \ref{fig:clustering}(b).

\begin{table}[htbp]
\caption{Moran's I values for spatially defined clusters}
\label{Tab:M_I_4clusters}
\centering
\small
\begin{tabular}{|c|ccc|}
\hline   
Spatial& \multicolumn{3}{c|}{Temporal Distribution}\\
\cline{2-4}
Cluster & Whole-week & Weekday & Weekend \\  
\hline         
   1 & \cellcolor{yellow}{\bf{0.64}} & 0.41 & 0.66 \\
   2 & 0.83	& \cellcolor{yellow}{\bf{0.80}} & \cellcolor{yellow}{\bf{0.84}} \\
   3 & 0.84 & \cellcolor{yellow}{\bf{0.81}} & \cellcolor{yellow}{\bf{0.81}} \\ 
   4 & \cellcolor{yellow}{\bf{0.63}} & 0.54 & 0.68 \\
\hline
\end{tabular}
\end{table}
The 2D temporal distribution of each cluster may be split into the weekday and weekend distributions (see section 3.2.1) if both of them have recognizable patterns and the two patterns are different. Table \ref{Tab:M_I_4clusters} lists the Moran's I values for the whole-week, weekday, and weekend distributions, respectively, for each cluster. The temporal distribution of cluster 1-weekday becomes more random compared to its whole-week distribution, which suggests to keep the whole-week distribution. Cluster 4 has the same situation. But the weekday and weekend distributions of both clusters 2 and 3 have strong temporal patterns. Furthermore, the Pearson's r between the weekday and the weekend distributions of cluster 3 is 0.34, indicating these two patterns are quite different. The Pearson's r between the weekday and weekend distributions of cluster 2 is 0.67, which means certain degrees of similarity and dissimilarity between these two patterns are both present. Therefore, this study decided to separate the weekday and weekend patterns for clusters 2 and 3. In total, six clusters of fatal crashes are determined. They are cluster 1, cluster 2-weekday, cluster 2-weekend, cluster 3-weekday, cluster 3-weekend, and cluster 4.

\subsubsection{Learning the Spatial-Temporal Attention} 

The spatio-temporal attention guidance for each of the six clusters highlighted in Table \ref{Tab:M_I_4clusters} was obtained through association rule mining. The rules provide the information about ``what type of fatal crash (in terms of FHE and MC) is likely to occur, and how likely, if a fatal crash would occur at the driving scene of such location and time."

For fatal crashes in each of the six clusters,  $x_{i,j}$ denotes the number of fatal crashes occurred at time spot $(i,j)$ (i.e., in month $i$ and hour $j$) and $x_{i,j,k}$ are those with type $k$ FHE. A rule that describes the association of type $k$ FHE with the time of crash is 
\begin{equation}
    x_{i,j}\Rightarrow x_{i,j,k}.
\label{eq:AR_ijk}
\end{equation}
There are many rules that can be extracted from the fatal crash data, and not all of them are important. The study computed the supports of $x_{i,j}$ and $x_{i,j,k}$, respectively, and the confidence and lift of each rule, to quantify the strength of the rule. Threshold values of these metrics are used to select meaningful rules to form the attention guidance.

For each of the six clusters, the support of $x_{i,j}$, denoted as $S_{i,j}$, calculates the fraction of fatal crashes occurred at time $(i,j)$:
\begin{equation}
    \text{Support}(x_{i,j}):=S_{i,j}=\frac{x_{i,j}}{\sum_{i,j} x_{i,j}}.
\end{equation}
$S_{i,j}$ helps determine if the time spot $(i,j)$ is worth attention for this cluster. 

Similarly, the support of $x_{i,j,k}$ calculates the fraction of fatal crashes with type $k$ FHE in this cluster:
\begin{equation}
    \text{Support}(x_{i,j,k}):=S_k=\frac{\sum_{i,j}x_{i,j,k}}{\sum_{i,j}x_{i,j}}.
\end{equation}
$S_k$ indicates if type $k$ FHE is worth attention for this cluster.

The confidence of the rule in Eq. (\ref{eq:AR_ijk}), denoted by $C_{i,j,k}$, calculates the likelihood that a fatal crash would have type $k$ FHE if it would occur at time $(i,j)$:  
\begin{equation}
    \text{Confidence}(x_{i,j}\Rightarrow x_{i,j,k}):=C_{i,j,k}=\frac{x_{i,j,k}}{x_{i,j}}.
\end{equation}
The confidence value measures the reliability of the rule. 

The lift of the rule in Eq. (\ref{eq:AR_ijk}), designated as $L_{i,j,k}$, measures the rise of the probability of type $k$ FHE with the time information being presenting over the same probability but without knowing the time:
\begin{equation}
    \text{Lift}(x_{i,j}\Rightarrow x_{i,j,k}):=L_{i,j,k}=\frac{C_{i,j,k}}{S_k}.
\end{equation}
The lift value measures the association between the time and the type of FHE.

The rules included as the attention guidance for any cluster must satisfy the following selection criteria: $S_{i,j}\geq 3.472\permil$ (the probability of the uniform distribution), $S_k\geq 5\%$, and $L_{i,j,k}>1$. MC related rules were discovered and evaluated using the same method and this paper skips the discussion.

\section{Discussion}
This study developed a rule query tool and the complete set of rules are provided at \cite{MATC_pics}. This section analyzes the rules, illustrates the implementation, and discussed the applicability.

\subsection{Spatio-temporal Attention Guidance with Visualization}

\begin{figure*}[htb]
    \centering
      \includegraphics[width=0.85\linewidth]{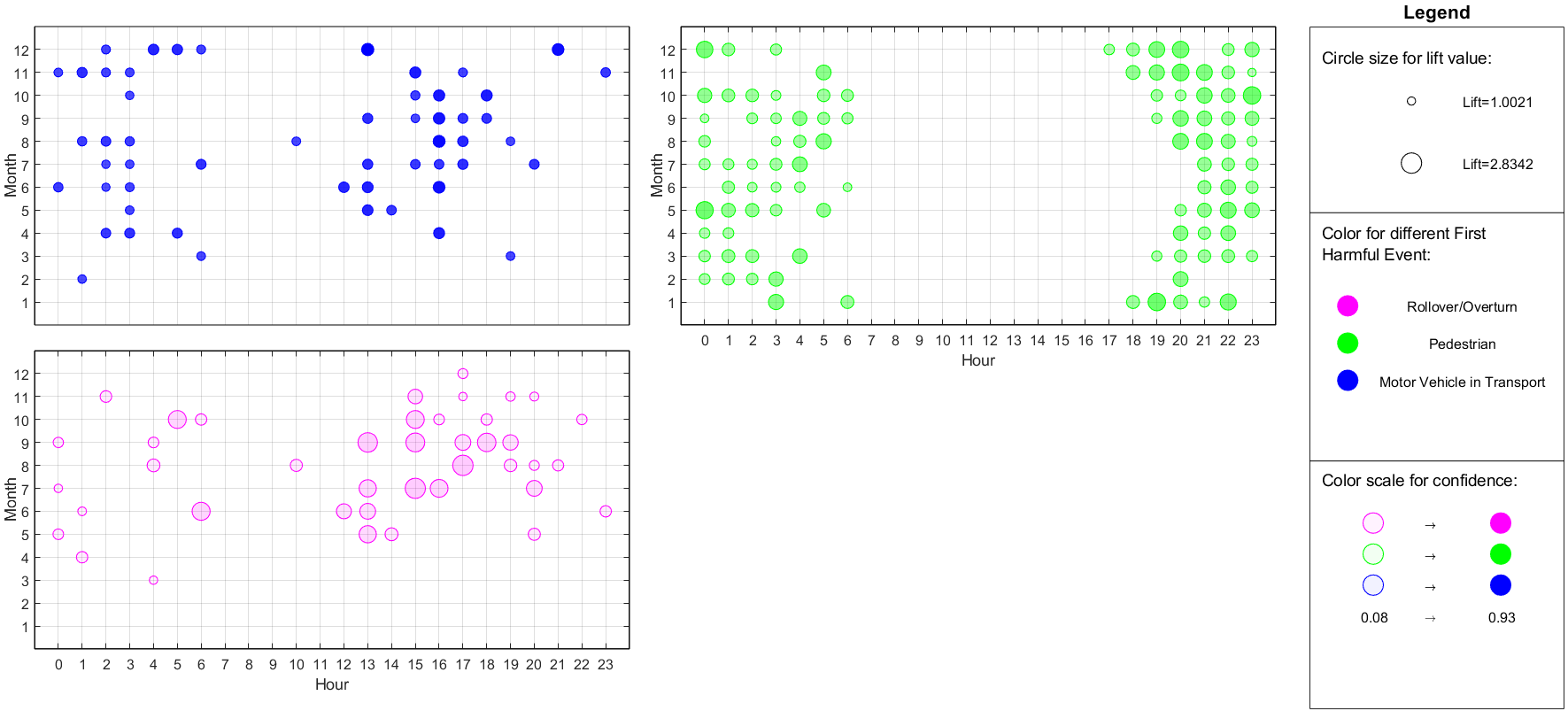}
    \caption{Association rules for analyzing the First Harmful Event (FHE) of fatal crashes in cluster 1}
    \label{fig:Harm_C1}
\end{figure*}
\begin{figure*}[htb]
    \centering
      \includegraphics[width=0.85\linewidth]{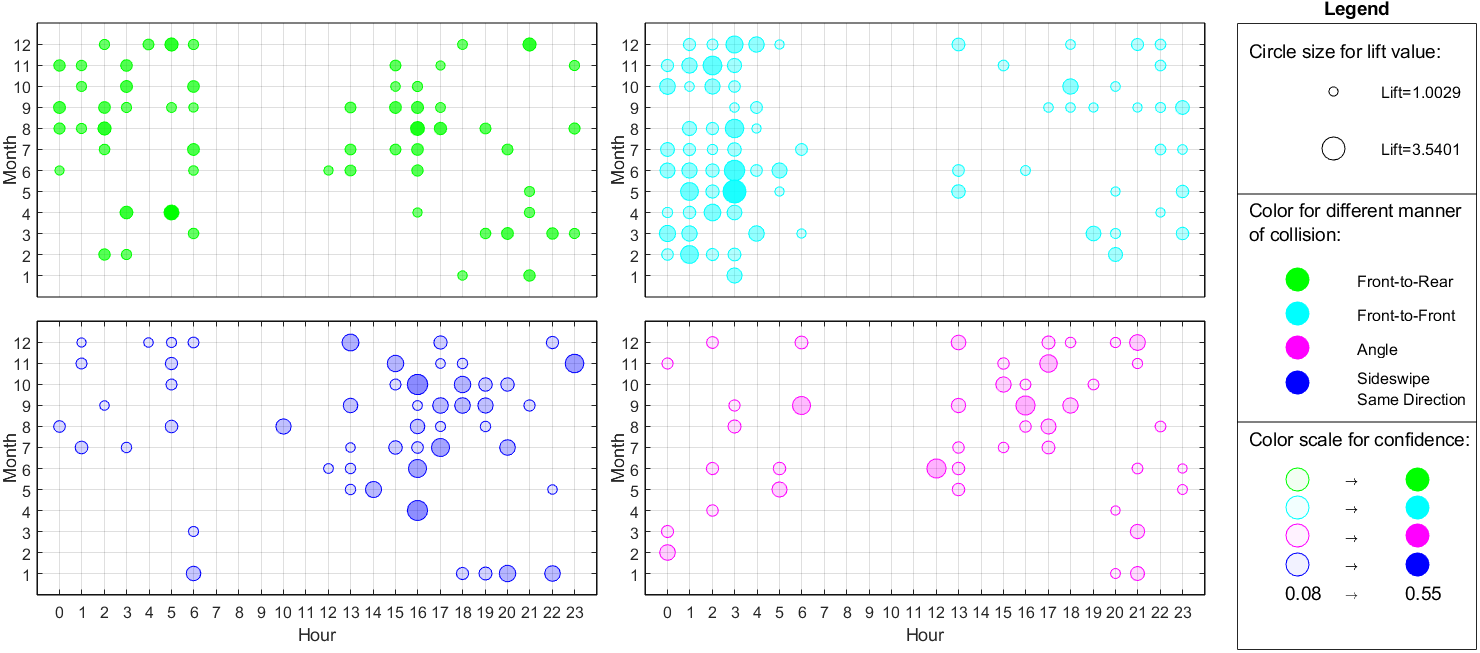}
    \caption{Association rules for analyzing the Manner of Collision (MC) of fatal crashes in cluster 1}
    \label{fig:Mancol_C1}
\end{figure*}
\begin{figure*}[hbtp]
    \centering
      \includegraphics[width=0.85\linewidth]{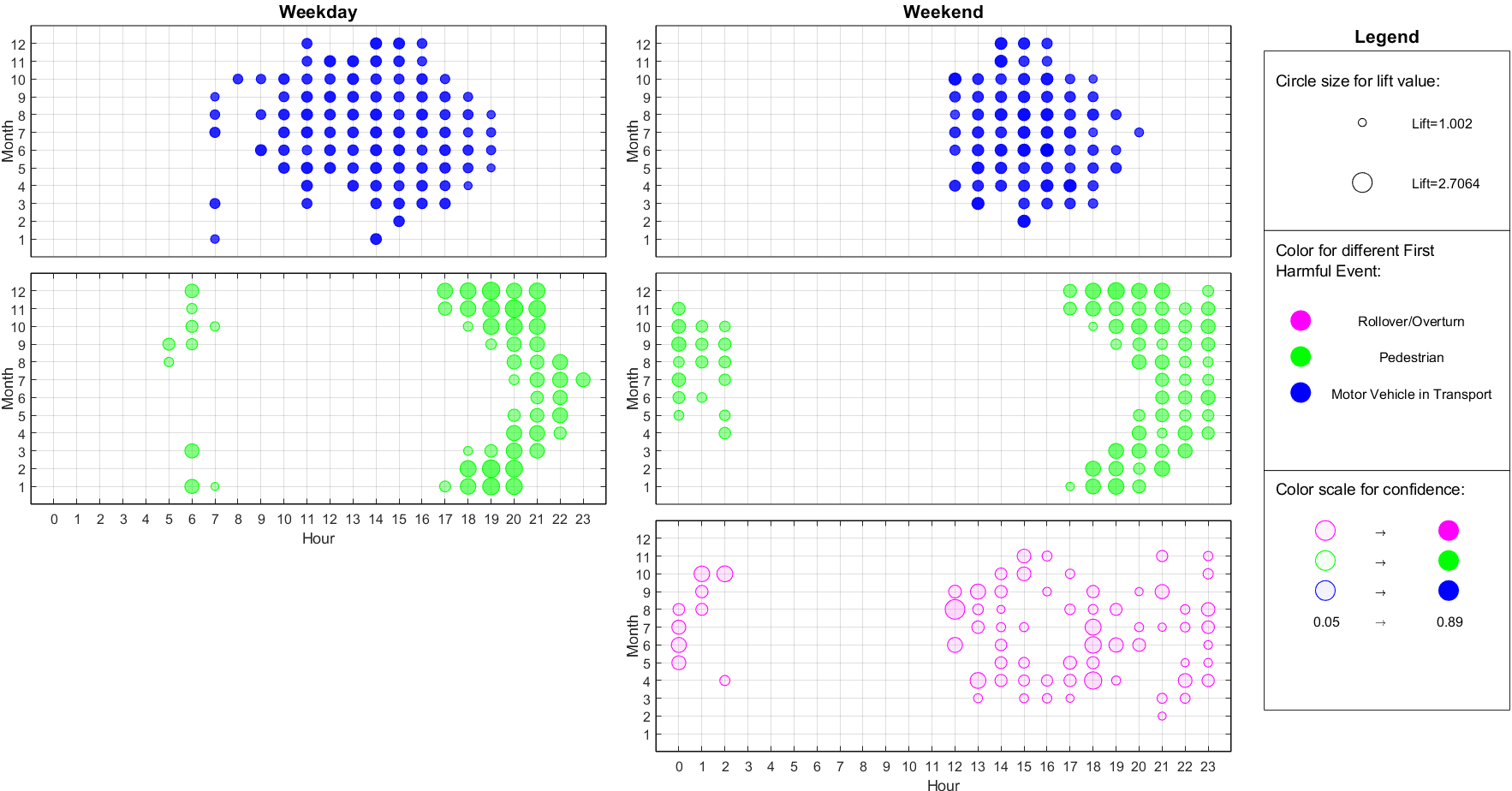}
    \caption{Association rules for analyzing FHE of fatal crashes in cluster 2 weekday/weekend}
    \label{fig:Harm_C2}
\end{figure*}

The spatio-temporal attention guidance for each of the six clusters is visualized to facilitate users' understanding. The paper discusses three examples below. Figure \ref{fig:Harm_C1} shows the attention guidance for cluster 1 with respect to FHE. Rules that meet the screening criteria defined in section 3.3.3 are only relevant to ``Motor Vehicle in Transport", ``Pedestrian", and ``Rollover/Overturn". Therefore, the guidance includes three month-hour grids with colored dots on the grids. Each grid is dedicated to one type of FHE. A dot on a grid represents an association rule for analyzing fatal crashes in locations of cluster 1 and at the time specified by the dot's position on the grid. The color of the dot indicates the FHE type, the color scale measures the confidence of the rule, and the size of the dot indicates the lift of the rule. The attention guidance is used in the following approach. For example, there are two rules that call for attention when driving at a location belonging to cluster 1 and at hour 0 of May. The first rule is the green dot at the intersection of May and hour 0 of the top right grid, which tells the following. If a fatal crash would occur, the chance that the FHE is ``Pedestrian" is 53.1\% (i.e., the confidence of the rule) and the probability of having the ``Pedestrian" FHE at this specific time is 2.2 times of the overall probability of having ``Pedestrian" FHE for cluster 1 (i.e., the lift of the rule). The second rule is the pink dot at the intersection of May and hour 0 of the bottom left grid, which is related to the ``Rollover/Overturn" FHE. This rule is weaker than the first rule. The confidence is 9.4\% and the lift is 1.2.

The three grids of dots in Figure \ref{fig:Harm_C1} are also three temporal patterns of fatal crashes split by the FHE type. The patterns convey useful information about fatal crashes in cluster 1. For example, fatal crashes with the ``Pedestrian" FHE mainly occurred at night, from 18:00 to 6:59. Fatal crashes with the ``Rollover/Overturn" FHE mainly occur from April to November. The overall confidence of the rules about the ``Motor Vehicle in Transport" FHE is the highest, followed by those with the ``Pedestrian" FHE and then ``Rollover/Overturn" FHE. Rules with large lift are mainly those with the ``Rollover/Overturn" FHE.

Figure \ref{fig:Harm_C1} shows many rules are about the ``Motor Vehicle in Transport" FHE. Therefore, rules about MC are further examined. Rules that meet the screening criteria are related to ``Front-to-Rear", ``Front-to-Front", ``Angle", and ``Sidewipe Same Direction".  Similarly, Figure \ref{fig:Mancol_C1} visualizes the MC related rules split by the MC type. The overall confidence of the rules about ``Front-to-Rear" MC is higher than that for other MC types. Rules related to the ``Front-to-Front" MC have larger variations in both lift and confidence, reflecting the heterogeneity of the rules in the month-hour temporal space. The same for rules related to the ``Sideswipe Same Direction" and the ``Angle" MC. Moreover, many fatal crashes with the ``Front-to-Front" MC are from midnight to the early morning.

Rules for analyzing the fatal crashes in cluster 2-weekday/weekend are displayed in Figure \ref{fig:Harm_C2}, which are split by the FHE type. The grid containing the rules of ``Rollover/Overturn" on weekdays is not included in Figure \ref{fig:Harm_C2} because no such rules meet the screening criteria specified in section 3.3.3. A comparison of Figure \ref{fig:Harm_C2} to Figure \ref{fig:Harm_C1} reveals the differences between clusters 2 and 1. Rules about the ``Motor Vehicle in Transport" FHE and those about ``Pedestrian FHE" a very small overlap on the month-hour temporal space. The attention guidance informs users to pay special attention to pedestrians at night (17:00-2:59) at the weekend, and at both night time (17:00-23:59) and morning rush hours (5:00-7:59) on weekdays. Users should also caution against fatal crashes with the  ``Motor Vehicle in Transport" FHE in the afternoon of the weekend (12:00-20:59) and during the daytime of weekdays (7:00-19:00). During the weekend, users should also caution against fatal crashes with the  ``Rollover/Overturn" FHE from 12:00 to 2:59 although rules of such fatal crashes are small in the confidence.

\subsection{Attention Guidance in Support of Visual Perception}

\begin{figure*}[htbp]
    \centering
    \includegraphics[width=0.9\linewidth]{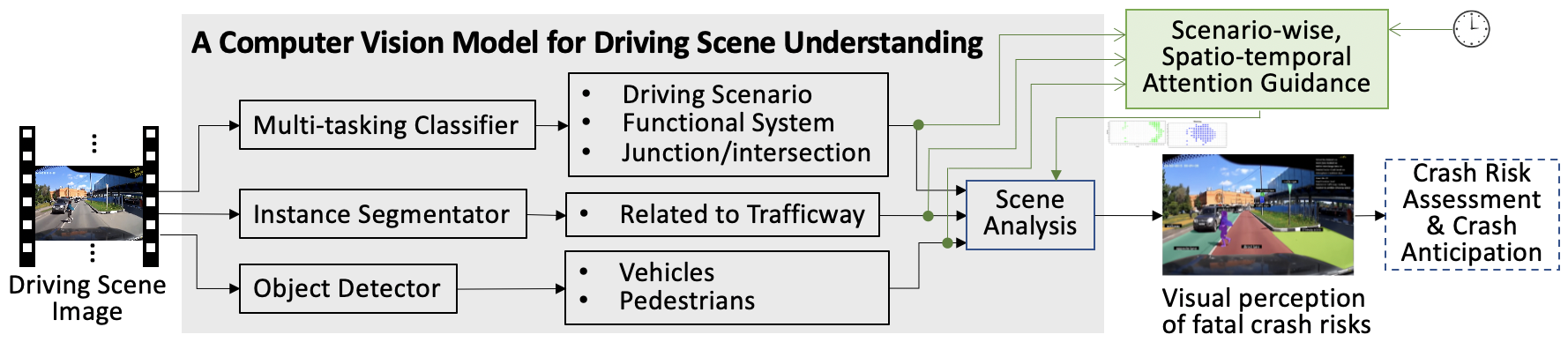}
    \caption{A computer vision model for visual perception of fatal crash risks}
    \label{fig:CVModel}
\end{figure*}

A Computer Vision (CV) model can be developed in association with the attention guidance to provide a preliminary analysis of driving scenes (RGB images) captured by the dashcam mounted on the user's vehicle. Figure \ref{fig:CVModel} shows that the CV model includes an image classifier, an instance segmentator, an object detector, and a scene analysis algorithm. First, a multi-tasking neural network is trained to classify input RGB images by the scenario, functional system, and junction/intersection. This network comprises of  multi-level perceptrons and a convolutional neural network.  Second, a deep neural network named DeepLab v3 is trained only for different drivable area segmentation related to trafficway (e.g., on the trafficway, on the roadside, and so on) with a road segmentation dataset. This deep network creates the pixel-wise partition of different areas related to trafficway. Third, an object detector named YOLO v3 is trained to detect concerning objects such as pedestrians and vehicles from the input images. YOLO v3 is adopted here instead of Deeplab v3, because Deeplab v3 is much more computationally expensive than YOLO v3 and YOLO v3 is advantageous on its bounding box generation. These parallel networks provide corresponding labels of the driving scene and they are used for two purposes. One purpose is to query and retrieve the relevant attention guidance for the scene, and the other purpose is to provide the information for the scene analysis that locates the detects objects and determines their relationships with the user's vehicle. Given the attention guidance as an input, the scene analysis provides the visual perception of fatal crashes and can better serve the crash risk assessment and crash anticipation in the near-crash phase. A preliminary version of this CV model has been developed by \cite{karim2021multi-task}. Interested readers may refer to it for the technical detail of the CV model, including the data for training, validating, and testing the model. 

\begin{figure*}[htbp]
    \centering
    \begin{subfigure}[b]{.49\textwidth}
    \centering
        \includegraphics[width=\textwidth]{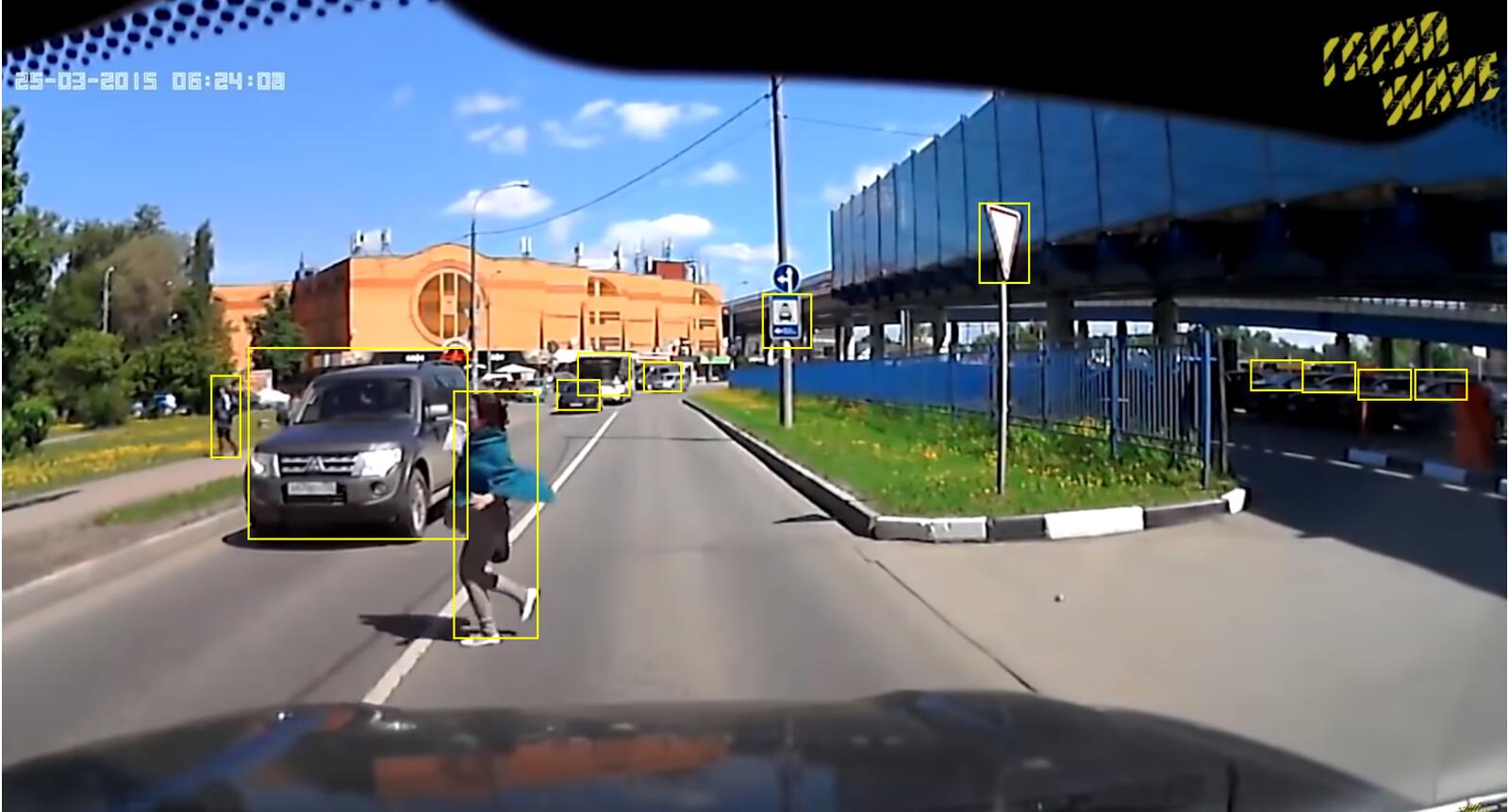}
        \caption{without attention guidance}
    \end{subfigure}
    \begin{subfigure}[b]{.49\textwidth}
    \centering
        \includegraphics[width=\textwidth]{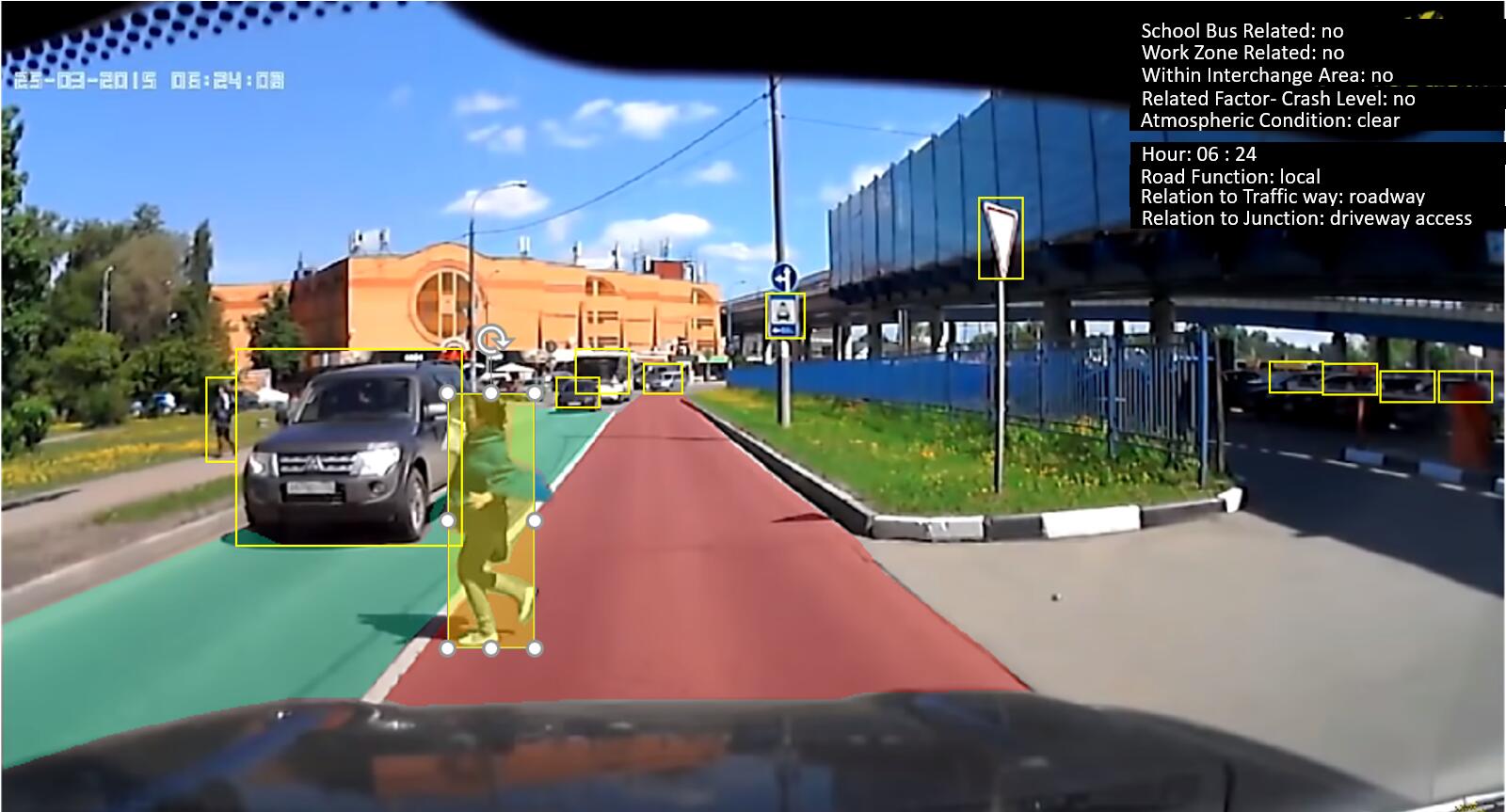}
        \caption{with attention guidance}
    \end{subfigure}
    \caption{An illustrative example of visual risk perception without and with the attention guidance}
    \label{fig:detailpic}
\end{figure*}

Figure \ref{fig:detailpic} is an example that illustrates the resulting difference of virtual risk perception by taking the scenario-wise, spatio-temporal attention guidance as an input to the scene analysis. Figure \ref{fig:detailpic}(a) presents the result without the attention guidance. Here, the network captures a variety of detected objects including vehicles and pedestrians. Without the attention guidance, the network cannot filter out the information irrelevant to fatal crashes from the scene. Moreover, the network passes all the information to the crash risk assessment and anticipation system, which not only complicates the assessment but may increase the chance of false alarm (i.e., predicting the occurrence of a crash that actually will not occur). Figure \ref{fig:detailpic}(b) presents the result with an input from the attention guidance. Details are discussed below.
\begin{figure*}[htb]
      \includegraphics[width=0.9\textwidth]{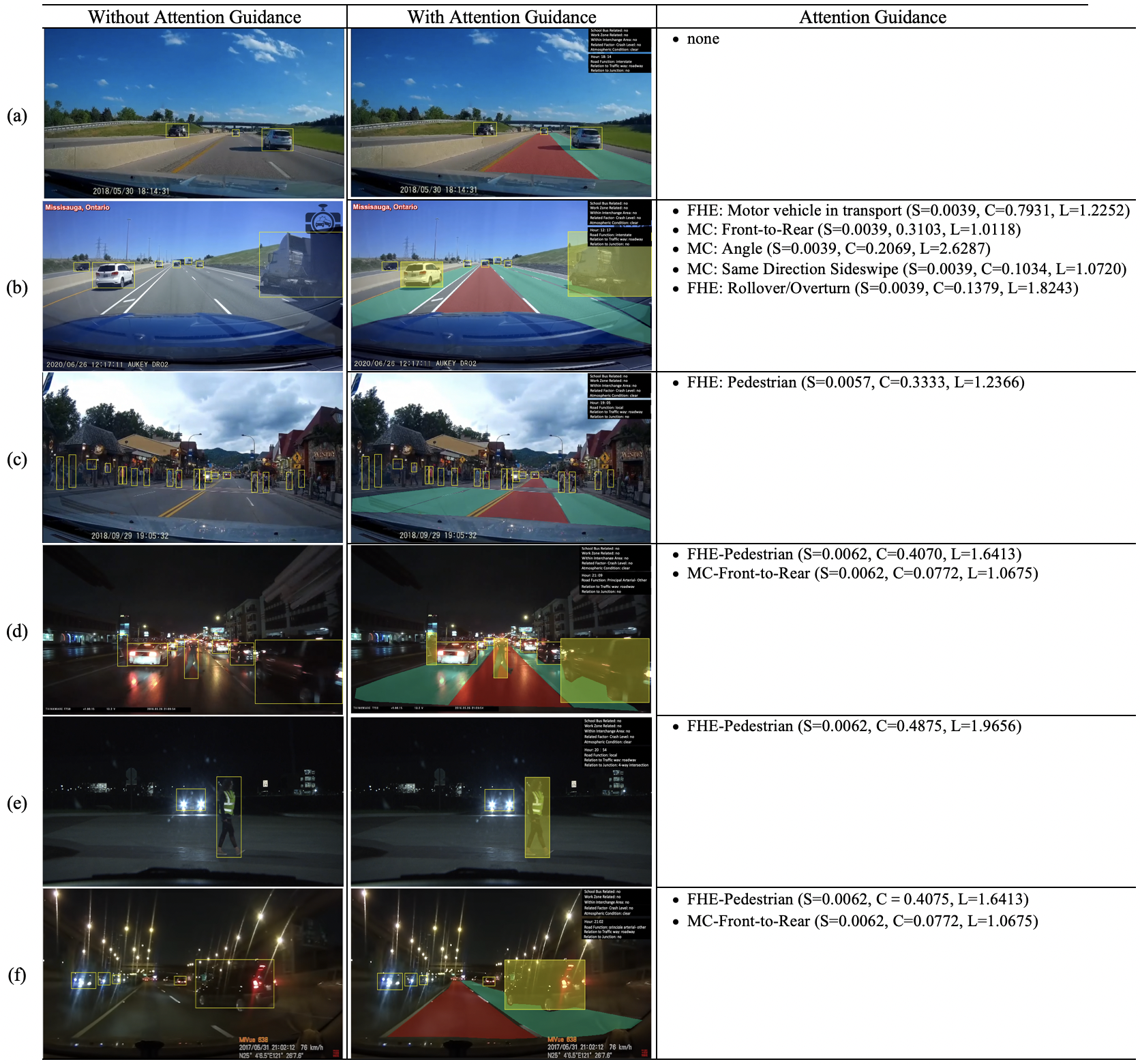}
    \caption{More examples of driving scene analysis with attention guidance (S, C, and L stands for $S_{i,j}$, $C_{i,j,k}$, and $L_{i,j,k}$ in Eqs. (5), (7), and (8))}
    \label{fig:9example}
\end{figure*}

{\bf Scenario Classification}. The CV model provides the labels for classifying crash scenarios (i.e. “School Bus Related”, “Work Zone”, “Relation to Junction-Within Interchange Area”, “Related Factor-Crash Level”, and “Atmospheric Conditions”), which are all zero. This means the driving scene belongs to the common scenario class. 

{\bf Spatio-temporal Information}. The CV model characterizes the spatial information of the driving scene according to the three location-related variables (i.e., ``FUN\_SYS", ``REL\_ROAD", and ``JUN\_INT"). Specifically, the classifier of the CV model determines that the road function class is ``Local" and the junction/intersection class is ``Driveway Access". The instance segmentator identifies different areas of the road including the direct lane, the opposite lane, and the walkway. Therefore, the location of the scene belongs to cluster 2. Moreover, the system clock tells the time is 6:24 am on a weekday in March. Given that the driving scene is in a location belonging to cluster 2, the chance of involving a fatal crash at 6:24am of weekdays in March is 3.7\permil.

{\bf Attention Guidance}. There are two rules that estimate the possible types of the fatal crash in case it would occur. The first rule is related to the ``Pedestrian" FHE with 0.4244 confidence  and 1.7114 lift.  This means the chance that the fatal crash has the ``Pedestrian" FHE is 42.44\% and this chance is 1.7114 times of the overall probability of ``Pedestrian" FHE at locations of cluster 2 and on weekdays. The second rule is related to the ``Front-to-Rear" MC with 0.0988 confidence and 1.3668 lift. It should be noted that rules about ``Motor Vehicle in Transport" FHE are not strong enough to be reported. Therefore, the rule of ``Front-to-Rear" MC here is a weak rule describing the likelihood of such collision if the fatal crash involves the ``Motor vehicle in Transport" FHE. 

{\bf Distance Measure.} The measurement of distance between the vehicles is adopted from our developed method in \cite{karim2021multi-task}.In this method, the input image will be divided into a 4X4 grid and concentrated on the area containing the travel path of the user's vehicle to reduce computational cost. Only vehicles in this highlighted region of interest are considered for the distance calculation. The distance between the vehicle in the image and the subject vehicle can be calculated by  the length of the dash camera $F$, the true height of the vehicle $H$, the height of the vehicle in the image $h$, where we assumes that $F$ is 2.5 inch, $H$ is 7 foot for a Van, 6 foot for an SUV, and 4.7 foot for a car, $h$ is measured in pixel. Same methodology has been applied to measure the distance of the pedestrian also.

{\bf Detection.} The method used for detecting drivable area and risky pedestrian is also adopted from \cite{karim2021multi-task}. DeepLab v3 trained with the BDD100k dataset \cite{yu2018bdd100k} was used in this study to segment the drivable area that includes two classes of lanes, the direct lane where the user is currently driving and the alternative lanes where the user can go by lane changing. DeepLab v3 cannot segment the area where objects like pedestrians and vehicles are present. Therefore, DeepLab by itself cannot identify the relationship of pedestrians to the driving lane. \cite{karim2021multi-task} proposed an algorithm of risky pedestrian detection. A rectangle bounding box of the segment is extrapolated using the most outside coordinates of the segmented direct lane. If the YOLO v3 object detector detects pedestrians in the driving scene, it returns all possible bounding boxes of pedestrians. If the bounding boxes of segment and pedestrian have an overlap, pedestrian is a risky one and a notification of risk is generated and the distance to that pedestrian is estimated.

{\bf Scene Analysis}. The two retrieved rules are used as the guidance for determining objects that are likely to involve in a fatal crash. The scene analysis algorithm of the CV model determines that the detected vehicles and pedestrians are in different locations. Some vehicles are outside the trafficway access (in the garage), some are very far away, and one vehicle is in the opposite lane (thus traveling in the opposite direction). One pedestrian is on the walkway along with the opposite lane and the other pedestrian is on the direct lane in front of the user's vehicle. The rule related to the ``Pedestrian" FHE suggests that the one in the direct lane probably deserves special attention but not the other pedestrian. The scene analysis algorithm calculates and finds that the distance to the pedestrian in front is about 5.1 ft, shorter than a pre-specified threshold 10 ft. Therefore, the pedestrian on the direct lane is labeled as a risky pedestrian and a risk notification makes the user aware of the risky pedestrian. Vehicles faraway and those outside the trafficway are certainly no longer considered for risk assessment. The weak rule related to the ``Front-to-Rear" MC suggests that vehicles driving in the same direction and in front of the user's vehicle may deserve  attention. Yet, the nearby vehicle in front is traveling in the opposite direction. Therefore, this vehicle is not likely to a fatal crash per the MC-related rule.

Six more examples of visual perception of fatal crash risks, (a)-(f), are listed in Figure \ref{fig:9example} to illustrate the support from the attention guidance. 
\begin{itemize}
\vspace{-0.8em}
\item[(a)] No rule is identified for scene (a) and thus no object needs special attention. But without the attention guidance, multiple detected vehicles, regardless of their locations, would be considered for further risk assessment.\vspace{-0.8em}
\item[(b)] The ``Motor vehicle in Transport" FHE is a strong rule for the scene. There are three additional rules describing the MC, including ``Front-to-Rear", ``Angle", and ``Sideswipe Same Direction". According to these rules, the two vehicles driving on the same direction and relatively close to the user's vehicle  are assigned the attention but not other vehicles. ``Rollover/Overturn" FHE is another rule, but it is less strong than the rule about ``Motor vehicle in Transport" FHE. Without these rules, additional vehicles that are traveling in the opposite direction or far away catch unnecessary attention. \vspace{-0.8em}
\item [(c)] A rule about the ``Pedestrian" FHE is retrieved for this scene. While many pedestrians are detected, the CV model can identify the locations of the pedestrians and the attention guidance suggests that those walking across the street deserve additional attention than than others.\vspace{-0.8em}
\item [(d)] Two rules are retrieved for this scene, one is related to the ``Pedestrian" FHE and the other is about ``Front-to-Rear" MC. Guided with the rules, two nearby pedestrians in front of the user's vehicle, one nearby vehicle traveling in the same direction, and one vehicle on the adjacent right lane are identified as objects that are more likely to involve in a fatal crash than others. \vspace{-0.8em}
\item [(e)] A rule related to the ``Pedestrian FHE" is retrieved, which suggests the pedestrian in front should be attended to, but not the vehicle driving across the intersection from the opposite direction.\vspace{-0.8em}
\item [(f)] Similar to scene (d) except no pedestrians are detected.\vspace{-0.8em}
\end{itemize}

\subsection{Applicability}
The scenario-wise, spatio-temporal attention guidance provides a method to discriminate the objects (e.g., vehicles, pedestrians, and others) detected from the dashcam footage according to the likelihood the objects probably involve in certain fatal crashes. Therefore, the result of the driving scene analysis supported by the attention guidance, if provided to drivers directly, has the potential to raise drivers' awareness about objects that deserve more attention for the safety purpose. The above-discussed examples have illustrated this functionality of the attention guidance. A rigorous assessment of the support to drivers is in process, which uses biometric sensors such as an eye movement tracker and an electroencephalogram in a lab setting to verify the change of attention and cognitive load of drivers due to the input from the attention guidance.

While the developed attention guidance can directly help drivers, the intention of creating this data analytics model is to support the design and implementation of a CV model that can perceive fatal crash risk earlier, before the near-crash phase. This capability can improve the learning of the spatio-temporal relationship of objects for crash anticipation during the near-crash phase. Crash prediction is still a challenging task of CV. The spatio-temporal relationship of the detected objects is critical information for the crash anticipation. \cite{chan2016anticipating} developed a Dynamic-Spatial-Attention (DSA) recurrent neural network for anticipating accidents from dashcam footage, which learns to distribute attention to candidate objects dynamically to gather subtle cues. \cite{bao2020uncertainty} highlighted the importance of the spatio-temporal relationship of accident-related agents in crash prediction. Introducing many unrelated objects to the relationship learning is likely to lower the accuracy of accident prediction. From this perspective, the attention guidance developed in this paper provides the information on highly relevant objects, backed by the historical data of fatal crashes. Research is taking place to integrate the early attention to crash-relevant objects with the crash prediction during the near-crash phase. 

The attention guidance estimates the relevance of an object to possible fatal crashes from the scenario class and the spatio-temporal information of the object. This approach is reliable for analyzing fatal crashes with sufficient records because the statistics of prior fatal crash records are the foundation for the relationship. It does not apply to those with small samples, for example, the ``Rollover/Overturn" FHE at locations of cluster 2 on weekdays.

\section{Conclusion and Future Work}
This paper presented a method to create scenario-wise, spatio-temporal attention guidance from fatal crash report data, which can support the design and implementation of a Computer Vision (CV) model for perceiving potential fatal crash risks from dashcam footage. With the attention guidance backed by big data of fatal crash reports, the CV model can estimate the relevance of detected objects to fatal crashes from their scenario classes and spatio-temporal information. The CV technique in stand-alone is difficult to achieve this capability early than the near-crash phase. Examples discussed in this paper have illustrated the resulting changes in the visual perception of fatal crash risks. The improved result would better support the crash risk assessment and crash anticipation in the near-crash phase.

Studies that support and extend this paper are from multiple aspects. This paper focuses on the methodological development for creating the attention guidance. While examples illustrate changes that the attention guidance can bring to the CV-based perception of fatal crash risks, rigorous validations of its direct support to human drivers and the main intention to improve the crash anticipation in the near-crash phase are important. For example, the use of biometric sensors such as an eye movement tracker and an electroencephalogram in a lab setting can verify the induced changes in focal attention and the cognitive load of drivers. The integration of the attention to crash-related objects early before the near-crash phase with the crash anticipation is not ready yet and requires extra research efforts. While this paper dedicates to the common scenario of fatal crashes, the knowledge of other scenarios is also important for reducing fatal crashes and thus needs thorough exploration. When new crash report data are released, the attention guidance needs to be updated to capture the long-term evolution of crash characteristics. This paper has built a solid foundation for exploring the above-mentioned future work.

\section*{Aknowledgement}
Financial support for this study is partially provided by the U.S. Department of Transportation, Office of the Assistant Secretary for Research and Technology (USDOT/OST-R) under Grant No. 69A3551747107 through the MATC University Transportation Center (http://ihttps://matc.unl.edu/). The views, opinions, findings and conclusions reflected in this publication are solely those of the authors and do not represent the official policy or position of the USDOT/OST-R, or any State or other entity.

\bibliographystyle{plain}
\bibliography{references}
\end{document}